\documentclass[titlepage,preprint,double-spaced]{article}
 \usepackage[cp1251]{inputenc}
\usepackage[russian]{babel} \usepackage{amssymb}
\usepackage{graphicx} 

\begin{document}
 \title {Magnetization and magnetostriction of
Van Vleck antiferromagnets with magnetic anisotropy of "easy-plane" $\:$ type}
\author{V.M.Kalita \thanks{Institute of physics of National Academy
of Science of Ukraine}, I.М. Ivanova \thanks{National Technical
University of Ukraine "KPI", Prosp. Pobedy 37, Kyiv 03056,
Ukraine}, V.M. Loktev \thanks {Bogolyubov Institute for
Theoretical Physics of National Academy of Science of Ukraine,
Metrolohichna Str 14, Kyiv 03680, Ukraine}}
 \maketitle

     The theoretical description of quantum phase transition, induced
     by the external magnetic field, into antiferromagnetic state in
     the Van Vleck -- singlet -- magnet with single-ion anisotropy of
     "easy-plane" $\:$ type and ion spin $S=1$ is proposed. It is
     shown that the spin polarization of the ground non-degenerated
     state proves to be the order parameter of such a transition and
     that the Landau thermodynamic approach can be employed for its
     (transition) description. The magnetic properties which include
     the field behavior of magnetization and magnetic susceptibility of
     the antiferromagnetic phase in the fields of different directions
     are studied. The peculiarities of induced magnetostriction in Van
     Vleck antiferromagnet, which as well as magnetization has a
     singularity in the phase transition point, are investigated. An
     attempt is made for qualitative comparison of results obtained
     with avaliable experimantal data.

    PACS number(s): 75.10.-b, 75.10.Jm, 75.30.Gw, 75.30.Kz,
    75.50.Ee

     $\,$
       \hrule

      \section{Introduction}
         It is well known, that magnetization of the classical (or, what is the
         same, weakly anisotropic) antiferromagnet at low temperatures
         (far below Neel one $T_N$), is connected with
         sublattice magnetizations turn only [1]. From this fact, it is usually supposed that
         their magnitude remains constant, and
          under the effect of the external magnetic field only
          their directions are changing. The character and peculiarities of
          magnetization process (spin-flip, spin-flop and also orientation phase transitions of the
           Ist kind)
          in such antiferromagnets depend on the next parameters: value and
          direction of the external magnetic field, anisotropy constants, and the magnitude of
          intersublattice exchange [2-5]. For example, in "easy-plane"
          $\:$ two-sublattice dihalogenids $\mathrm{NiCl_2}$ or
          $\mathrm{CoCl_2}$ of iron group field bahavior of magnetization [6-10] is well
          satisfied by quasi-classical approach, although these
          magnets defer significantly. The value of "easy-plane" $\:$ single-ion
          anisotropy in $\mathrm{NiCl_2}$ is much less than exchange -- each ion orbital moment
          in crystal field is practically  fully frozen. At the same time there is
          only partial freezing of orbital moment in $\mathrm{CoCl_2}$, and the
          "easy-plane" $\:$ single-ion anisotropy is approximately the half of exchange
           (by order of its value [7]). The field behavior of induced magnetostriction of these
          crystals [11-13] are also agreed with idea of rotation of
          preserving by module of sublattice magnetizations.

             Among antiferromagnets there are, however, some family of crystalls, in which
             single-ion anisotropy exceeds inter-ion exchange [14,15] -- it is
             so-called Van Vleck, or singlet, antiferromagnets. The magnetic ordering in them
             is absent at all temperatures, up to $T=0$. Such materials
             include, in particular, hexagonal crystals of $ABX_3$ type, where $A$
             is the ion of alkali metal ($A=\mathrm{Cs,Rb}$), $B$ is the
             transition metal ($B=\mathrm{Fe}$), $X$ is the halogenyde ($X=\mathrm{Cl,Br}$). In these crystals
             magnetic moments, induced by external field on paramagnetic ions $B^{2+}$, form
             antiferromagnetic chains along $C_3$ axis, on the one hand, and, on the other, -- triangular
             structures in basic plane (see reviews [16-19]).
             There are also some other compounds, that refer to
             Van-Vleck antiferromagnets, and among them so-called DTN,
             which chemical formula is $\mathrm{NiCl_2}$-$\mathrm{4SC(NH_2)_2}$ [20-23]. It
             also has (antiferromagnetic) chains Ni-Cl-Cl-Ni along
            "heavy" $\:$ magnetic axis, although in field absence the
            mean
             spin on each site is equal to zero, because of exceeding by single-ion
             anisotropy of parameters of both intra- and
             intersublattice exchange. It should be emphasized, that DTN
             can be refered to the group of two-sublattice Van Vleck
             antiferromagnets, that have different from $\mathrm{NiCl_2}$
             crystal structure, another character of exchange
             interactions, which by the value are much less  than
             single-ion anisotropy [22,23].

        The magnetization process in Van Vleck antiferromagnets differs principally from
        that which takes place in classical Neel antiferromagnets
        [25-28]. At first, the magnetic ordering without external magnetic field is
        absent  and therefore there are
        no any magnetic sublattices. Secondly, magnetic, or particularly
        antiferromagnetic, ordering in Van Vleck antiferromagnets can
        appear spontaneously by quantum (in definition of Ref. [24]) phase
        transition, induced by magnetic field [16-23].

     Weak dependence of magnetic susceptibility on external magnetic field,
     thus, represents some sort of peculiarity of anriferromagnetic
     phase. As a result, the observed magnetization actually follows
     the linear field behavior [22,23,29,30]. It appears, in other
     words, that such a behavior of magnetization of a system (but not
     the proper sublattice magnetization) in antiferromagnetic phase is
     similar to the magnetization induced by external field in Neel
     antiferromagnets. It could be understood, if the transition to the
     antiferromagnetic phase would be the phase transition of the Ist
     kind. In such a case sublattices could magnetize, in the
     transition point, due to jump (in the presence of corresponding
     susceptibility singularity), and at further field growth there can
     occur sublattice magnetic vectors turn only. The experiment shows,
     however, that transformation of non-magnetic (singlet) state in
     the antiferromagnetic one takes place continuously, what means,
     that this magnetic transformation is the phase transition of IInd
     kind [22,23,29,30]. The latter demonstrates, that sublattice
     magnetizations appear and change also continuously from their
     initial zero value to maximal one. So, the classical approach with
     the constant module of sublattice average spin for Van Vleck
     antiferromagnets is not applicable fundamentally.

     The results of DTN induced magnetostriction measurements are
     presented in the papers [22,23]. It could be seen from them,
     particularly, that induced magnetostriction in DTN appears and
     exists namely in the antiferromagnetic phase only. There was also found,
     that relative crystal deformation, that arised along "heavy" $\:$
     magnetic axis, changes its sign during field increasing.$\,$ Such a
     behavior of DTN striction was attributed in Ref. [23] to the
     prevailing role in this compound of intersublattice magnetoelastic
     interaction of exchange nature [20,21]. On the other hand, the
     magnetostriction sign change is also observed in some classical
     Neel antiferromagnets, for example in $\mathrm{CoCl_2}$ [11-13],
     where it is conditioned by the anisotropic intrasublattice
     magnetoelastic interaction.

     Thus, the description of induced magnetostriction in the Van Vleck
     antiferromagnets, where anisotropy is not small, and so,
     anisotropic magnetoelastic interaction also can be comparable (or even larger) with
     isotropic exchange one, becomes relevant. The
     consideration of this problem requires to account the fact, that
     sublattice magnetization modules in antiferromagnetic phase of the
     initially singlet magnet essentially depend upon the field.

     From the above said it can be evident, that there are some
     unresolved questions of the description of induced magnetic phase
     transition into the antiferromagnetic phase and its magnetic
     characteristics (field dependencies of sublattice and system as a
     whole magnetizations, magnetic susceptibility, magnetostriction) in Van
     Vleck systems.

     Below we shall proceed from the suggestion, that in Van Vleck
     magnets the intrinsic spontaneous magnetic (or antiferromagnetic)
     moment is equal to zero, so for them there is no magnetic ordering
     temperature without the external magnetic field. It would seem,
     that the absence of magnetic (dipole) moment , or, in other words,
     magnetization (spin) on the site, shows not only the absence of
     any magnetic ordering, but, clearly, the absence of the magnetic
     contributions in physical properties of corresponding systems.
     However, in fact, this is not right, because the absence of
     ordinary -- exchange-induced -- spin ordering do  not include the
     presence the ordering of other type in, particularly, the
     quadrupole one. The latter, in one or other way, is peculiar to
     all Van Vleck magnets, which are the special case of magnetic
     crystals with more specific -- nematic -- type of spin ordering
    [31,32].

     The one or another spin ordering proves itself not only by the
     appearance in crystal of new (spin-)electron excitation branches,
     which, for example, in Van Vleck nematics turn out gapped. It can
     also be revealed in such an observed and calculated characteristic
     of these magnets as their magnetostriction, which peculiarities
     for such systems is not completely studied yet. A the same time,
     because of recent calculations of DTN magnetoelastic features
     [22,23], there is definitely such a need.

     It can also be noted, that in some papers (for example, Refs. [33-36]),
     the description of phase transition between singlet and induced
     atiferromagnetic states carries out by using the representation of
     Bose-Einstein condensation of magnons. Indeed, the appearance of
     magnetization in finite magnetic fields can be formally described
     in the terms of some magnetic excitations condensation. But in
     reality in observed systems there does not occur any true
     condensation of quasiparticles, because, as it will further be
     seen, one should say about rearrangement of the ground state only,
     and thus -- about virtual, but not real magnons [37].

     Below there is considered the model of strongly anisotropic,
     two-sublattice antiferromagnet with ion bare spin $S=1$. In the
     framework of quantum approach it will be made an attempt to
     describe the crystal magnetization, magnetic susceptibility and
     magnetostriction at magnetically induced phase transition from the
     initial singlet state to the spin-ordered one. For calculation of
     physical characteristics of a system, there will be used the total
     energy $E$, that is the sum of relevant contributions:

      \begin{equation}
      E=E_{exch}+E_{an}+E_{h}+E_{el}+E_{m-el},
      \end{equation}
     where $E_{exch}$ is the exchange energy; $E_{an}$ is the magnetic
     anisotropy energy; $E_{h}$ is Zeeman energy; $E_{el}$ is the
     elastic energy and $E_{m-el}$ is the magnetoelastic energy, or the
     energy of spin-lattice coupling. It is also supposed, that
     magnetoelastic interactions are much weaker then exchange ones,
     and do not have a noticeable feed-back influence on the magnetic ordering.
     Assumptions made allow to provide a calculation in the simplest,
     but rather general form, confining, as usual, by summands, that
     are linear by elastic deformation tensor in the magnetoelastic
     energy and are quadratic by this tensor in elastic energy. So, at
     this limitations, there can be solved a problem of the magnetic
     ordering, at first, and only then use obtained results for field
     dependence of induced striction.

      \section{The ground state of singlet antiferromagnet with $S=1$
    and "easy-plane" $\;$ magnetic anisotropy in the longitudinal magnetic field}

  In accordance with above mentioned the consideration will be for simplicity restricted by bilineal anisotropic
  (intra- and intersublattice) exchange interactions, single-ion "easy-plane" $\;$ anisotropy
  and Zeeman term. The simplest model Hamiltonian of a system, that defines
  three contributions, $E_{exch}$, $E_{an}$ and $E_{h}$, in Eq. (1), can be written as:

     \begin{eqnarray}
    H=\frac{1}{2}\sum_{\mathbf{n}_{\alpha}, \mathbf{m}_{\beta}}
    J_{\mathbf{n}_{\alpha}\mathbf{m}_{\beta}}\mathbf{S}_{\mathbf{n}_{\alpha}}
    \mathbf{S}_{\mathbf{m}_{\beta}}+\frac{1}{2}\sum_{\mathbf{n}_{\alpha}, \mathbf{m}_{\beta}}
    J^{Z}_{\mathbf{n}_{\alpha}\mathbf{m}_{\beta}}S^{Z}_{\mathbf{n}_{\alpha}}
    S^{Z}_{\mathbf{m}_{\beta}}
     +D\sum_{\mathbf{n}_{\alpha}}\left(
     S^{Z}_{\mathbf{n}_{\alpha}}
   \right)^2-h_{\parallel}\sum_{\mathbf{n}_{\alpha}}S^{Z}_{\mathbf{n}_{\alpha}},
         \end{eqnarray}
    where $\alpha,\,\beta=1,2$ are the magnetic sublattice indices,
    which numbers in the considered system was chosen as 2; vectors
    $\mathbf{n}$ and $\mathbf{m}$ gives spins position in magnetic
    sublattices, which are described
    by spin operators $\mathbf{S}_{\mathbf{n}\alpha}$;
    the constant $D>0$, that reflects an "easy-plane" $\;$ magnetic
    structure; the magnetic field $h$ is defined in units of energy, so
    $h_{\parallel}=\mu_BgH_Z$; $H_Z$ is the OZ projection of
    magnetic field, at that the crystallographic symmetry axis OZ
    is perpendicular to the "easy" $\,$ plane. Just at the $\mathbf{h}\parallel OZ$
    the magnetic field induce the phase transition to the
    antiferromagnetic state. The case of transverse field $\mathbf{h}\perp
    OZ$ will also be considered below, but it should be emphasized, that
    such a field do not induce any phase transitions. Parameters
    $J_{\mathbf{n}_{\alpha}\mathbf{m}_{\beta}}$ characterize the value of exchange interaction
    isotropic part and $J^Z_{\mathbf{n}_{\alpha}
    \mathbf{m}_{\beta}}$ is exchange anisotropy, which can
    be basically both "easy-axis" $\;$ and "easy-plane" $\;$ types.
    We, however, will assume, that inter-ion anisotropy, as also
    single-ion one, relates to the same -- $\;$"easy-plane"$\;$ -- type.

    The convenience of these restrictions is conditioned by the fact, that in such
    a physical
    situation both sublattices become symmetric relatively to the
    external field, what allows to reduce twice the number of
    equation derived.

    The analysis of possible quantum eigenstates of Hamiltonian (2)
    at $\mathbf{h}\parallel OZ$ will be provided, using the
    approximation of self-consistent field, that corresponds to spin
    fluctuation neglecting and to the change of average by
    multiplying spin operators of different sites on multiplying of
    averages. In this case the energy $E_{gr}$ of the ground state,
    normalizing on one cell (for nearest both inter- and intrasublattice different spins)
    is equal to:

     \begin{eqnarray}
     E_{gr}=\frac{1}{2}\sum_{\alpha\beta}J_{\alpha\beta}z_{\alpha\beta}\mathbf{s}_{\alpha}\mathbf{s}_{\beta}
     +\frac{1}{2}\sum_{\alpha\beta}J_{\alpha\beta}z_{\alpha\beta}s^{Z}_{\alpha}s^{Z}_{\beta}
    +D\sum_{\alpha}Q^{ZZ}_{\alpha}-h_{\parallel}\sum_{\alpha}s^{Z}_{\alpha}
     \end{eqnarray}
     where $\mathbf{s}_{\alpha}$ is the quantum-mechanical average
     of spin vector of $\alpha$th sublattice in the ground ion state;
     $z_{\alpha\beta}$ is the number of nearest neighbors from the same ($z_{\alpha\alpha}$)
     and another ($z_{\alpha\beta}\equiv z_{12}$) sublattices.
     Also here are introduced such averages for components
     of spin quadrupole moment $Q^{ZZ}_{\alpha}$ [40-42]. It should
     be also noted that for antiferromagnet the intersublattice
     exchange parameter is $J_{12}z_{12}\equiv I>0$;  at the same
     time, the parameter $J_{11}z_{11}=J_{22}z_{22}\equiv J$ of
     intrasublattice exchange can be of any sign, that we will also
     choose for simplicity as furthering to the ordering, $J<0$. The
     exchange anisotropy, in this case, satisfies the conditions of
     its "easy-plane" $\;$ type: $J^Z_{12}z_{12}\equiv \triangle I<0$ and
     $J^Z_{11}z_{11}=J^Z_{22}z_{22} \equiv \triangle J>0$.

      Let us impose for spins of each sublattice  their
     proper (rotating) coordinate systems
      $\xi_{\alpha}\eta_{\alpha}\zeta_{\alpha}$,
       that $\alpha$th sublattice average spin
      is always oriented along $\zeta_{\alpha}$ axis,
      what means that this axis is
      the quantization one for this spin sublattice,
      and $\xi_{\alpha}$ axis is lain in $Z\zeta_{\alpha}$
      plane. Thus, in such coordinate systems the correct wave
      function of the $\alpha$th sublattice ground spin state, as
      well known, has next form [38,39]:

         \begin{equation}
     \psi^{\left(
     0\right)}_{\alpha}=\cos\phi_{\alpha}\mid1>+\sin\phi_{\alpha}\mid{-1}>,
     \end{equation}
     where $\mid\pm 1>$ and $\mid 0>$ are eigenfunctions of operator
     $S^{\zeta}_{\mathbf{n}_{\alpha}}$ in bra-ket representation.
     Next, it can be calculated, using (4) the quantum-mechanical
     spin and quadrupole averages:

     \begin{eqnarray}
      s=\cos2\phi,\quad Q^{\zeta\zeta}=1,\quad
     Q^{\xi\xi}=\frac{1}{2}\left( 1+\sin2\phi\right).
     \end{eqnarray}

      In the expressions (5) the sublattice indices
      are missed, because as  mentioned at the chosen field direction
      the evident dependence of observables on index
      $\alpha$ is absent.

      The usage of functions (4) allows one to describe the energy (3)
      at $\mathbf{h}\parallel OZ$ as:

     \begin{eqnarray}
     E_{gr}=I\cos^22\phi\cos2\theta-|J|\cos^22\phi-J_Z\cos^22\phi\cos^2\theta
     \nonumber\\+2D\left[ \cos^2\theta+\frac{\sin^2\theta}{2}\left(
     1+\sin2\phi\right)\right]-2h_{\parallel}\cos\theta\cos2\phi,
     \end{eqnarray}
     where $J_Z\equiv \triangle J-\triangle I$.

    For the determination of magnetization, magnetic
    susceptibility and subsequently striction, there should be found field
    behavior
     of mean spin (and its direction) for each sublattice in the field.
     Also it should be made the same calculation for
      spin quadrupole moment. As it was reported in
     Refs. [40,41], the solution of the problem of spin configuration
     in the magnetic field consists of the minimization of the
     expression (6) by all available unknown quantities: the geometrical angle
     $\theta$ and (see Eq. (4)) the angle $\phi$ of quantum
     states mixture. Such a method of observables finding,
    being completely an equivalent to the solution
     of quantum self-consistent problem, is more convenient and consistent, because
     allows the generalization on the case of finite temperatures
     [27,28].

    The equations for both required angles are:
     \begin{eqnarray}
     \frac{\partial E_{gr}}{\partial \phi}=-2\left(
     I\cos2\theta-|J|+J_Z\cos^2\theta\right)\sin4\phi
     +2D\sin^2\theta\cos2\phi
     +4h_{\parallel}\cos\theta\sin2\phi=0,\\
     \frac{\partial E_{gr}}{\partial \theta}=-\left(
     2I+J_Z\right)\sin2\theta\cos^22\phi
     -D\sin2\theta\left(
     1-\sin2\phi\right)+2h_{\parallel}\sin
     \theta\cos2\phi=0.
     \end{eqnarray}

     It is known from Ref. [37], that the set of Eqs. (7)-(8) in the
     absence of the external magnetic field has two solutions: the
     non-magnetic one, $s=0$, that exists at $D>2(I+|J|)$ and the
     "magnetic"$\;$ one at $D\leq2(I+|J|)$, with which the reduced
     value of single-site mean spin

     \begin{equation}
     s=\sqrt{1-\frac{D^2}{4\left( I+|J| \right)^2}}<1
     \end{equation}
     is associated.

    The initial ground state of the system should be the singlet
    one,
      $s=0$, that the quantum phase transition (at the
    magnetic field $\mathbf{h}\parallel OZ$) from this state to
    magnetically ordered one occurs. So, below we will suppose, that the next
    evident inequality $2(I+|J|)/D<1$ takes place. At this  model parameters
    ratio, the ground state of the system is really nonmagnetic and the
    ordering in the absence of magnetic field is not realized at any
    temperatures [41]. Otherwise, this ratio actually gives the
    condition on singletness of magnet ground state, which is Van
     Vleck one. The solution $s=0$  also satisfies Eqs. (7)-(8) for some interval of
     magnetic fields.

     As field grows the finite value, $s\neq0$, of the mean spin on the site
     appears. It can be derived from Eq. (8) the
     expression for the average spin orientation relatively to the
     crystallographic axis:

     \begin{equation}
     \cos\theta=\frac{h_{\parallel}\cos2\phi}{D\left( 1-\sin2\phi\right)+
     \left( 2I+J_Z\right)\cos^22\phi}.
     \end{equation}

      It is seen from Eqs. (7) and (10), that in large fields, when $h_{\parallel}\geqq
      h_{flip}$ (where $h_{flip}\equiv D+2I+J_Z$) the state, in
      which spins of both sublattices are directed along "heavy"$\;$
      ($\theta=0$) axis, is established. Then the spin projecton on the
      external field direction will be maximum and equal
      to $s=S=1$. For $h_{\parallel}<h_{flip}$ spins of sublattices
      are orientated at finite angle $0<\theta\leq\pi/2$ to the
      "heavy"$\;$ axis.

          \section{Thermodynamic analysis of spin states}
   Using the formulas (5) and substituting Eq. (10) into the Eq. (6), it could be
   obtained the ground state energy in the form of functional
   \begin{eqnarray}
   E_{gr}=-\left( I+|J|\right)s^2+D\left(
   1-\sqrt{1-s^2}\right)
  -\frac{h^2s^2}{D\left(
   1+\sqrt{1-s^2}\right)+\left( 2I+J_Z\right)s^2},
   \end{eqnarray}
     which depends on the spin polarization $\mathbf{s}$ only.
    The expansion of this energy over the small $s$ gives:
   \begin{equation}
   E_{gr}=\frac{h_s}{D}\left(
   h_s-h_{\parallel}\right)s^2+\frac{D}{8}\left( 1+
   \frac{2h^2_s\left( 2I+J_Z\right)}{D^3}\right)s^4
     \end{equation}
    where $h_s=D\sqrt{1-2(I+|J|)/D}$ is the critical field of
    magnetization appearance.

    In the expansion (12), which refers to the field region $h_{\parallel}\rightarrow
    h_s$, one can restrict by terms, that are not higher then of 4th
    power by $s$. Actually this expansion for the ground state energy
    is similar to the free energy expression in Landau theory of
    phase transitions. However, in Eq. (12) the ground state spin polarization
    corresponds to the order parameter, and the leading parameter,
    that results in the phase transition, is not the temperature,
    but the magnetic field. It can be seen from Eq. (12), that at
    $h_{\parallel}<h_s$ the coefficients at
    $s^2$ and $s^4$ are positive, and so the ground state of spin
    system will be Van Vleck non-magnetic single-ion state.
    At the pint
    $h_{\parallel}=h_s$ the sign of coefficient at $s^2$ changes,
    and in the fields $h_{\parallel}>h_s$ the spin
    polarization (of still non-degenerate ground state)
    spontaneously appears. The value of polarization can be readily found by
    minimization of $E_{gr}$ (12):

         \begin{eqnarray}
     \frac{\partial E_{gr}}{\partial s}
     =2s\left[ \frac{h_s}{D}\left(
     h_s-h_{\parallel}\right)+\frac{D}{4}\left( 1+\frac{2h^2_s\left(
     2I+J_Z\right)}{D^3}\right)s^2
     \right]=0.
     \end{eqnarray}

     From Eq. (13) it follows, that near the critical point
     $h_{\parallel}\geq h_s$ this polarization
     (or simply the spin of the ground state)
     fundamentally depends on the field:

     \begin{equation}
     s\left( h_{\parallel}\right)=\sqrt{\frac{4h_s\left( h_{\parallel}-h_s\right)}
     {D^2+2h^2_s\left( 2I+J_Z\right)/D}}.
     \end{equation}

     In the same vicinity,  $h_{\parallel}\geq h_s$, of the critical
     point the angle $\theta$ between vector $\mathbf{s}$
     and axis $OZ$ is determined by the expression:

     \begin{equation}
     \cos\theta=\frac{h_s}{2D}\sqrt{\frac{4h_s\left( h_{\parallel}-h_s\right)}
     {D^2+2h^2_s\left( 2I+J_Z\right)/D}}.
     \end{equation}

     Thus, it is found, at $h_{\parallel}=h_s$ the spin
     polarization spontaneously arises as field grows in the
     very "easy"$\,$ plane, because at $h_{\parallel}-h_s\rightarrow 0$
     the angle $\theta \rightarrow \pi/2$. In other
     words, it turns out, that in the moment of its appearance,
     the vector $\mathbf{s}(h_{\parallel}\geq h_s)$ is
     perpendicular to the longitudinal field: $\mathbf{s}\perp\mathbf{H}\parallel OZ$.
     Further magnetic field growth leads not only to the
     decreasing of $\theta$, at it follows from Eqs. (14)-(15),
     but also to the simultaneous increasing of spin polarization, that is as
      bigger its value, as more it flatten
     against the "heavy"$\,$ axis.

     In the whole, the induced tilt of the magnetic
     subalttices, and thereafter the magnetization of Van Vleck
     antiferromagnet include two processes: the classical rotation
     of spins (sublattice magnetizations) and
     purely quantum (because of angle $\phi$ change) growth of
     single-site polarization $s(h_{\parallel})$. Both
     processes take place also at $T=0$. The antiferromagnet magnetization
     (normalizing on one magnetic atom) is described by the evident product:

     \begin{equation}
     m_{\parallel}\equiv m\left( h_{\parallel}\right)=s\left(
     h_{\parallel}\right)\cos\theta=\frac{2h^2_s\left( h_{\parallel}-h_s\right)}
     {D^3+2h^2_s\left(2I+J_Z\right)}
     \end{equation}

     As a result, one arrives to the unexpected result: the observed
     magnetization near critical field of quantum transition from singlet to
     spin-polarized state depends linearally -- as in classical
     antiferromagnets -- upon the external magnetic
     field, that induce the very transition. From this comes another rather
      remarkable
     conclusion: at such a phase transition the
     magnetic susceptibility of a system should have a jump.

     Thus, in the framework of approach, that is similar
     to the Landau thermodynamic one, it was
     demonstrated, that the spin polarization is the only order parameter for
    induced by magnetic field $\mathbf{h}\parallel OZ$ quantum phase transition from
      Van Vleck phase to the antiferromagnetic one.
       But despite the fact, that
      calculation was made for the case $T=0$, the
      required polarization proves to be essentially dependent
      on the external field. It should be reminded,
      that in classical antiferromagnets ion spin polarization is
      fixed at $T=0$ and do not evaluate in the field, while in Van
      Vleck system it appears as a consequence (in terminology of Ref.
      [24]) of quantum phase transition [27,28].

      Next an attention should be drowned to such an analogy, that
      single-ion anisotropy, reducing average spin, plays a role
      of "disordering" $\;$ factor, and in this sense can be
      compared with entropy. It (single-ion anisotropy) leads to
      the mixture (or linear combination of quantum states), that results in the
      absence of spin polarization of ions in their ground state.
      The exchange and magnetic fields, on the contrary,
     resist to this, "magnetizing" $\,$ the system
      and causing the spontaneous (or forced) spin polarization,
      which at the moment of its appearance is directed
      perpendicularly to the magnetic field.

      The studied quantum phase transition between Van Vleck (also ordered,
      in essential) and antiferromagnetic states is, as it was
      seen, the consequence of different interactions (exchange, Zeeman and
      spin-orbital, that
      lies at the heart of single-ion anisotropy) competition.
      Therefore such a quantum transformation is natural to
      identify as the magnetic phase transition of "displacement"$\,$ but not of
       "order-disorder"$\,$ type. As distinct from the last one, the
      transition of displacement type is not the transition in the system of
      spins, which fluctuate between degenerated (or almost
      degenerated) quantum states "up" $\,$ and "down", because the
      ground state of quantum paramagnets is always non-degenerated and
      its polarization is the direct consequence of this state
      rearrangement in the external field.

      There should be noted, at last, that the applicability
      of phenomenological theory, that is based on the expansion
      (11), is confined by the fields
      $h_{\parallel}\geq h_s$ in the vicinity of critical point
      $h_s$. In the field region $h_{\parallel}>>h_s$ the
      magnetization process should be analyzed with more
      exact expressions both for ground state energy and for the
      equations, which define the spin configurations. However the latter can be
      easily found numerically.

  \section{The magnetization curves and magnetic susceptibility
of antiferromagnetic phase}

 Substituting in the Eq. (11) the expression (5)
  the equation, that describes the spin polarization
 as the function of longitudinal field, can be
 obtained:

 \begin{eqnarray}
 s\left( D-2\left(I+|J| \right)\sqrt{1-s^2}-\frac{D\left(1+\sqrt{1-s^2}
 \right)h^2_{\parallel}}{\left(D\left( 1+\sqrt{1-s^2}\right)+\left( 2I+J_Z\right)
  \right)^2}\right)=0.
  \end{eqnarray}

  It should be noted, that this equation refers both the fields $h_{\parallel}<h_s$
 of the existence of Van Vleck phase, where
  (see Eq. (12)) $h_s=\sqrt{D^2-2D(I+|J|)}$, the nonmagnetic,
  $s=0$, state is stable, and to the region $h_s\leq h_{\parallel}\leq
  h_{flip}$ of the antiferromagnetic phase (up till the
  point $h_{flip}$ of its flipping). It is obviously, that at
  the point (see Eq. (10)) $h_{\parallel}=h_{flip}$, which is
  corresponded to the value $\theta=0$, the polarization arrives to its maximum value
  $s=1$ on the site.

  Using Eq. (17), the behavior $s(h_{\parallel})$
   in the fields $h_s\leq h_{\parallel}\leq h_{flip}$ can be found, and
  from Eq. (10) -- also angle $\theta$. After that, it is not difficult
  to define the field dependence of the quadrupole $Q^{ZZ}_{\alpha}$ in Eq.
  (3). In the framework of such an approach the field dependencies of
  magnetization were calculated (see Fig.1).

  \begin{figure} \includegraphics[width=0.7 \textwidth]{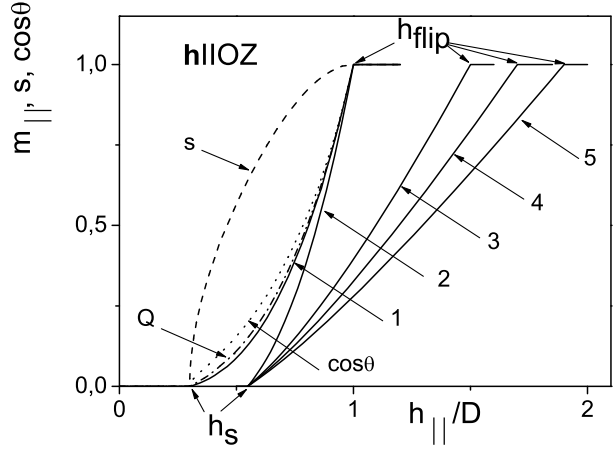}
  \caption{Magnetization $m_{\parallel}$ (solid curves),
  spin $s$, quadrupole moment $Q$ and value of $\cos \theta$ \it versus \rm field.
  The curves for $s$, $Q$ and $\cos\theta$ are calculated at $|J|/D=0.455$ and
  at condition $I=J_Z=0$. Magnetization lines 1-5 are for next parameters:
  line 1 is for  $|J|/D=0.455$, $I=J_Z=0$; line 2 is for
  $|J|/D=0.35$, $I=J_Z=0$; line 3 is for
  $|J|/D=0.35$, $I=0$, $J_Z/D=0.5$; line 4 is for  $|J|/D=0$, $I/D=0.35$, $J_Z=0$;
  line 5 is for $|J|/D=0.15$, $I/D=0.2$, $J_Z/D=0.5$.}
  \end{figure}

     The curve 1 on Fig.1 was plotted for such case, when
  intrasublattice exchange is prevailing, while intersubalttice
  and anisotropic ones are omitted. At chosen
  parameters the magnetic sublattices are fixed artificially, because this
  extreme case corresponds actually to the two independent
  antiferromagnets, The field dependencies for $s$, $Q^{ZZ}_{\alpha}\equiv Q$ and
  $\cos\theta$ are also shown on  Fig.1 for these parameters.
 . It can be seen, that at the point $h_{\parallel}=h_s$ the average site spin
  really spontaneously appears, and exists in the region
  $h_{\parallel}>h_s$. With further field growth, the value of $s$
  is increasing, leading by velocity of the change of angle $\theta$.
  This velocity, however, becomes more fast, when the
  field approaches to the flipping field and, correspondingly,
  $s \rightarrow 1$. In such a case the intrasublattice exchange (due to its
  isotropy) has no effect on spin saturation, and so the
  value $h_{flip}$ of this critical field is completely defined by the
  single-ion anisotropy.

  From curve 1, that refers to the case $D-2|J|<<D$, it follows,
  that the fields $h_{flip}$ and $h_s$ differ quite weakly
   ($h_{flip}/h_s\approx 3$), although in the experiment [22,23]
   their ratio reaches 6, and from the data of Ref. [30] this
   ratio is about 8. Besides, the field dependence of
   magnetization for the considered case $I=0$ reveals, as can be
   seen, large nonlinearity, while the experimental data for all
   above mentioned compounds, are evidence of near linear
   dependence of magnetization on magnetizing force.

   It should be noted, that the case, in which inequality $(D-2|J|)/D<<1$ is satisfied,
   is physically available, but it can not be justified from the
   experimental point of view. To demonstrate this on Fig. 1 the
   curve 2 is plotted, for which the difference
   between parameters in intrasublattice exchange and single-ion
   anisotropy is chosen not less, but bigger than for curve 1.
   This choice really leads to the increase of the field value
   $h_s$ and do the decrease of the ratio $h_{flip}/h_s$, what
   indicates, that in attempts of interpretation of the experimantal
   magnetization, the intersublattice exchange can not
   be neglected.

   It is interesting, that when $I/D \rightarrow 0$, Eq. (17) has
   an exact solution, using which the ground state energy can be
   written in the form of function of magnetic field:

 \begin{equation}
 E_{gr}=\frac{1}{4D^2|J|}\left( h^2_{\parallel}-h^2_s\right)^2.
 \end{equation}

Then the magnetization (normalizing on one magnetic ion again) takes next form:
 \begin{equation}
 m_{\parallel}=-\frac{\partial E_{gr}}{\partial
 h_{\parallel}}=\frac{h_{\parallel}}{2D^2|J|}\left(
 h^2_{\parallel}-h^2_s \right).
 \end{equation}

 The dependence (19) is described by the lines 1 and 2 on Fig. 1.
 From Eqs. (18) and (19) it can be also seen quite a big field
 nonlinearity of magnetization in the antiferromagnetic phase. The
 expression (19), for fields $h_{\parallel}\rightarrow
 h_s$ can be also presented in the form of Eq. (16), when $2|J|\rightarrow D$.

 Now let consider the opposite limiting case, when intersublattice exchange
  is the biggest one in the system.
 It is seen, that even if one preserves the
 exchange field (which formally gives the same value of $h_s$),
 which affects on the spin from other sublattice, the
 change of magnetization (curve 4 on Fig. 1) occurs. The
 antiferromagnetic (intersublattice) exchange, unlike
 the intrasublattice one, leads to the growth of the field $h_{flip}$,
 because in this case the external field should overcome the effect of the
 same anisotropy, from one side, and, from the other side, -- of
 exchange field, that prevents parallel orientation of sublattice spins.

 The curve 3 already shows the
 nonlinearity decrease in $m(h_{\parallel})$, which as
 if it is rectified by intersublattice exchange (or by its anisotropic part).
 At the same time, the antiferromagnetic exchange together with
 the external magnetic field (in the region
 $h_{\parallel}>h_s$), resisting the anisotropy, leads to the
 establishment of spontaneous polarization. In the large fields,
 when polarization tends to its maximum value, the behavior of
 exchange in Van Vleck antiferromagnet does not differ from that
 one in classical antiferromagnets: it simply resists to the parallel
 configuration of both sublattice spins.

Curves 3 and 5 demonstrate the influence of easy-plane exchange anisotropy.
Actually this anisotropy does not change the position of critical field $h_s$, but it
also does not "want"$\,$ the establishment of collinear state, when $\mathbf{s}_1\rightarrow
\mathbf{s}_2\parallel OZ$. At the same time the account of exchange anisotropy of
 easy-plane type allows to obtain such a behavior of
 magnetization, that is near to linear one and observed in Refs.
 [22,23,29,30]. For clarifying how good the linear dependence
 corresponds to
 $m(h_{\parallel})$, there is shown on Fig. 2 the magnetic
 susceptibility $\chi_{\parallel}=dm_{\parallel}/dh$ for the same
 parameters as on the Fig. 1.

 Because the magnetization is nothing other, then
 $m_{\parallel}=s\cos\theta$, where $s=s(h_{\parallel})$, the
 longitudinal magnetic susceptibility of Van Vleck magnets is naturally
 to represent in the form of
 two above mentioned terms -- the classical $\chi_{cl}$ and the quantum
 $\chi_{quan}$ ones, so that
  \begin{eqnarray}
 \chi_{\parallel}=\chi_{cl}+\chi_{quan};\qquad\nonumber\\
 \chi_{cl}=s\sin\theta\frac{\partial \theta}{\partial h};\quad
 \chi_{quan}=\cos\theta\frac{\partial s}{\partial h}.
     \end{eqnarray}

   As can be seen from Fig. 1, near $h_s$ the biggest growth
   reveals
   the spin polarization $s$, so in the fields $h_{\parallel}\rightarrow
   h_s$ the "quantum" $\,$ contribution will dominate in the  magnetic
   susceptibility. And when the value of spin polarization
   saturates ($s(h_{\parallel}\rightarrow h_{flip})\rightarrow
   1$), the susceptibility will be mainly controlled by
   classical term (see Eq. (20)).

      \begin{figure} \includegraphics[width=0.7 \textwidth]{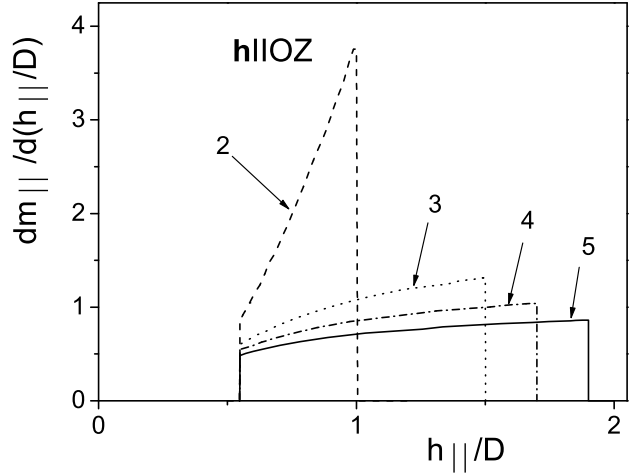}
   \caption{Magnetic susceptibility
   $\chi_{\parallel}$ \it versus \rm field.
      The numbers of curves are corresponded with parameters, that were used for
   the plotting of lines with the same numbers, that are on Fig. 1}
   \end{figure}

      Figure 2 shows, that when intrasublattice exchange is really
   largest one, then magnetic susceptibility grows, changing in
   field region $h_s\leq h_{\parallel}\leq h_{flip}$ in four
   times. If intersublattice exchange and/or exchange anisotropy
   "switches"$\,$ on, then field dependence of differential magnetic
   susceptibility $\chi_{\parallel}\equiv \chi(h_{\parallel})$
   becomes essentially weaker. Nevertheless, it can be seen from
   plots, shown on Fig. 1, that nonlinearity of function
   $m(h_{\parallel})$ in the antiferromagnetic phase
     at the chosen parameters remains quite noticeable.

     The case, when within the boundaries of this phase the value
     of $\chi(h_{\parallel})$ changes (30-50\%), is shown on Fig. 3,
     which meets the model parameters $|J|/D=0.05$, $I/D=0.3$, $J_Z/D=1$ or $J_Z/D=1.5$. In other words, the exchange
     anisotropy is comparable or even exceeds the single-ion one.
     At such ratios between the parameters the field $h_{flip}$ is
     almost in five times exceeds the field $h_s$ (one should note, that
      experimentally observed ratio is $h_{flip}/h_s\approx 6$
      [22,23]).

      On the same Fig. 3  the dependencies of $s(h_{\parallel})$, $\cos\theta(h_{\parallel})$
      and $Q(h_{\parallel})$ are shown for parameters  $|J|/D=0.05$, $I/D=0.3$ and $J_Z/D=1$.
      The behavior of $Q(h_{\parallel})$ almost
      coincides (see Fig. 1) with the field dependence $m(h_{\parallel})$.
      Moreover, it follows from Fig. 3, that exchange anisotropy, even a comparable
      with single-ion one, do not fully linearize the function
      $m(h_{\parallel})$. As was mentioned, this fact can be explained
      by the existence of two different regions in the
      magnetization of Van Vleck magnet.

        \begin{figure} \includegraphics[width=0.7 \textwidth]{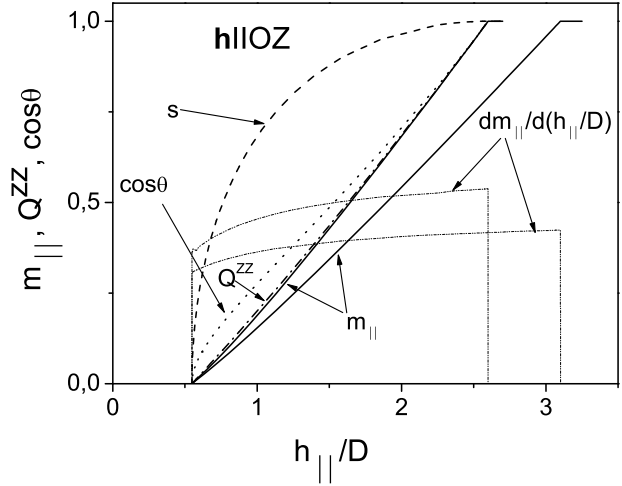}
      \caption{Longitudinal magnetization $m_{\parallel}$
      and magnetic susceptibility $\chi_{\parallel}$ \it versus \rm
      field at $|J|/D=0.05$, $I/D=0.3$,
      $J_Z/D=1$ and 1.5. The functions $s(h_{\parallel})$,
       $\cos\theta(h_{\parallel})$
      and $Q(h_{\parallel})$ are shown only for $|J|/D=0.05$,
      $I/D=0.3$, $J_Z/D=1$.}
      \end{figure}

      On the first of these regions, near $h_s$, the quantum
      process, as was pointed out, is determinant and the magnetization is
      defined basically by the appearance and growth of $s(h_{\parallel})$.
      On the second one, in the vicinity of $h_{\parallel}\leq
      h_{flip}$, the more important becomes the classical
      rotation of sublattice spins to the field direction, at
      essentially less (but not absent) role the very spin value
      change. It is obvious, that in this
      region the susceptibility depends much weaker on the value of
      magnetic field. So, it can be supposed, that the flipping of
      tilted antiferromagnetic sublattices, or the
      transition of Van Vleck system between induced
      two- and one-sublattice magnetically ordered states occurs as
      an orientation phase transition in ordinary antiferromagnet, when
      the variation of sublattice magnetization directions is
     in fact the only process.

       However, even for this field transition the quasiclassical
       approach do not give the correct solution for
       $m(h_{\parallel})$ in spin nematic. Indeed, one could
       suppose, that near the flipping field, when $s(h_{\parallel}\rightarrow
       h_{flip})\approx 1$, the quasiclassical magnetic energy in the
       ground state takes the form:

  \begin{equation}
  E_{gr}=I\cos2\theta+\left(
  J_Z+D\right)\cos^2\theta-2h_{\parallel}\cos\theta=0.
  \end{equation}
   Then from Eq. (21) immediately follows the equation

  \begin{equation}
  \frac{dE_{gr}}{dh_{\parallel}}=2\left[ -\left( 2I+J_Z+D\right)\cos\theta+
  h_{\parallel}\right]\sin\theta=0,
  \end{equation}
  which shows, that in the vicinity of $h_{\parallel}\rightarrow
  h_{flip}$ the magnetization of tilted ($\theta \neq 0$) phase is
  proportional to the field:
  $m_{\parallel}=\tilde{\chi}_{\parallel}h_{\parallel}$, where
  \begin{equation}
  \tilde{\chi}_{\parallel}=\frac{1}{D+2I+J_Z}\equiv
  \frac{1}{h_{flip}}=const.
  \end{equation}

  It can be seen, that in the field $h_{flip}$,
  the magnetization (on one spin) is $m_{\parallel}=1$. However, the
  susceptibility (23) differs from the exact ratio (20) and
  gives physically incorrect behavior of magnetization. It is connected with the fact,
  that
  in the region $h_{\parallel}\rightarrow h_{flip}$ it appears,
  that $m_{\parallel}$ depends linearly versus
   magnetic field, and asymptotically tends to zero at
  $h_{\parallel}\rightarrow 0$. As for plots, shown on Figs. 1 and
  3, it is easy to see, that the function $m(h_{\parallel})$,
  although it behaves linearly by field, but nevertheless it
  depends on magnetic field in not a direct proportion.

  An important conclusion follows from this: the quasiclassical
  approach (21), based on the substitution of quadrupole moment $Q^{ZZ}$
  by the average spin Zth projection square, appears to be
  unapplicable even in such a field region, where spin polarization
  almost reaches its saturation value $s\rightarrow 1$.

\section{Field behavior of magnetization in transversal field}

As it was reminded, the phase transition to the antiferromagnetic state does not occur at
$\mathbf{h}\perp OZ$, although the magnetic field magnetizes the
system.

Lets suppose, that due to the antiferromagnetic exchange in the easy plane,
two sublattices are formed. Then their spins
lay in this plane and are identically tilted to the field.
 In this case the energy of the ground state is:

 \begin{eqnarray}
 E_{gr}=I\cos2\varphi\cos^22\phi-|J|\cos^22\phi+D\left(
 1+\sin2\phi\right)-2h_{\perp}\cos\varphi\cos2\phi,
 \end{eqnarray}
where $\varphi$ is the angle between vector $\mathbf{s}_1$ (or vector $\mathbf{s}_2$)
and field $\mathbf{h}$, and the angle between $\mathbf{s}_1$ and $\mathbf{s}_2$ is twice
larger, $2\varphi$.

The spin configuration will be defined, as always, by minimizing the energy (24).
As a result, it is the set of equations (cp. Eqs. (7) and (8)):

  \begin{eqnarray}
 \frac{\partial E_{gr}}{\partial
 \varphi}=-2I\cos^22\phi\sin2\varphi+2h_{\perp}\sin\varphi\cos2\phi.
 \end{eqnarray}
 \begin{eqnarray}
 \frac{\partial E_{gr}}{\partial \phi}=-2\left(
 I\cos2\phi-|J|\right)\sin4\phi+2D\cos2\phi+4h_{\perp}\cos\varphi\sin2\phi=0
 \end{eqnarray}

 The equation (25) has two solutions. For the first of them
 $\varphi=0$ and it corresponds to one-sublattice magnetization,
 when the polarization of magnetic ions is directed along the
 field. The second solution $\cos\varphi=h_{\perp}/(2I\cos 2\phi)$
 provides the existence of two sublattices. The last, taking into
 account Eq. (5), can be rewritten in the usual form:

 \begin{equation}
 \cos\varphi=h_{\perp}/2Is.
 \end{equation}

 The denominator of Eq. (27) is the intersublattice exchange field,
  and this expression is similar to the
 expression for the magnetic sublattice tilt angle in classical
 antiferromagnets [1,2]. Nevertheless,it should be noted, that
 spin in Eq. (27) is not equal to its maximum value.

 Substituting (27) in (26), one arrives to the equation:

 \begin{equation}
 2\left[ 2\left( I+|J|\right)\sin2\phi+D\right]\cos2\phi=0
 \end{equation}

It follows from Eq. (28), that the spin polarization for antiferromagnetic
state at $\mathbf{h}\perp OZ$ should be equal to (9). However, the model
parameters, accepted above, are such, that the denominator under the root in Eq. (9)
is larger than 1, and non-polarized singlet is the ground state of ions.
Thus, the solution (27) is possible only for initially polarized
 ground state,
or when the antiferromagnetic (not singlet)
 phase is realized in the system even at $h_{\perp}=0$. But if without field
 the polarization  is $s=0$, then from the set
of Eqs. (25)-(26) fundamentally another result follows: the critical field
of polarization appearance in the transversal geometry is the $h_{\perp}=0$.
The distinction from "longitudinal"$\,$ case, for which the critical field is
finite, is easy to explain. At any fields $h_{\perp}\neq 0$ the ground state
with $S_Z=0$
(in crystal coordinate system) is immediately admixed  with the ionic exited states,
which have $S_Z=\pm 1$. In the case of longitudinal field, there is a threshold
for such an admixture.
  Then taking into account, that the transversal
 field does not induce the antiferromagnetic phase, one can
 obtain:

   \begin{figure} \includegraphics[width=0.7 \textwidth]{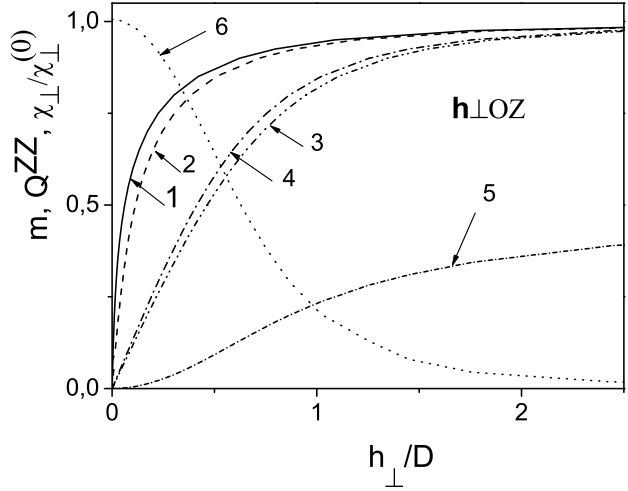}
 \caption{Field behavior of $m(h_{\perp})$ (lines 1-4),
 $Q^{ZZ}(h_{\perp})$ (line 5) and $\chi_{\perp}$ for $\mathbf{h}\perp
 OZ$. The line 1 is calculated for parameters $|J|/D=0.455$, $I=0$,
 the line 2 is for $|J|/D=0.35$, $I=0$, the line 3 is for $|J|/D=0$, $I=0.35$,
 lines 4-6 are for $|J|/D=0.05$, $I=0.3$}
 \end{figure}

 \begin{equation}
 E_{gr}=\left(
 I-|J|\right)s^2+D\left(1-\sqrt{1-s^2}\right)-2h_{\perp}s.
 \end{equation}
 It should be also noted, that at $\mathbf{h}\perp OZ$ the
 spin polarization is always equal to magnetization, which is
 directed along $\mathbf{h}$, i.e. $m_{\perp}=s$. Minimizing
 energy
 (29), one readily arrives to the equation

 \begin{equation}
 \frac{\partial E_{gr}}{\partial s}=2\left(
 I-|J|\right)s+D\frac{s}{\sqrt{1-s^2}}-2h_{\perp}=0,
 \end{equation}
 that allows to define the dependence of spin polarization upon
 the transversal field.

 On Fig. 4 there is shown the field behavior for $m_{\perp}$,
 that are obtained from Eq. (30). It is seen,
 that if intrasublattice exchange dominates in the system, then the
 magnetization increases rapidly, and if intersublattice exchange
 is $\,$"added", then the magnetization slows down.

 Despite the fact, that at $\mathbf{h}\perp OZ$
 the average spins are oriented also perpendicularly to Z,
 the spin quadrupole moment $Q^{ZZ}$ versus field
 reveals the behavior (see Fig.4) similar to the spin polarization.
  The magnetization rate decreases as the field grows and the magnetic
 susceptibility has a maximum at $h_{\perp}\rightarrow 0$.
 The normalized magnetic
 susceptibility $\chi^{(0)}_{\perp}=\chi(h_{\perp}=0)$ is also plotted on Fig. 4.

 Using Eq. (30), the expression for magnetic susceptibility at
 $\mathbf{h}\perp OZ$ can be obtained; it has the form:

 \begin{equation}
 \chi_{\perp}\equiv \chi\left( h_{\perp}\right)=\frac{1}{2\left( I-|J|\right)
 +D\left( 1-s^2\right)^{-3/2}}
 \end{equation}
It is seen, that in large fields, when $s\rightarrow 1$,
 transversal susceptibility $\chi_{\perp}\rightarrow 0$.
 In the opposite limit $h_{\perp}\rightarrow 0$, the magnetic
susceptibility is equal to:

 \begin{equation}
 \chi^{\left( 0\right)}_{\perp}=\frac{2}{D+2\left(I-|J|\right)}
 \end{equation}

It also follows from Eq. (32), that when  the
intrasublattice exchange increases the value of $\chi^{(0)}_{\perp}$ grows and, on
the contrary, at the increasing of intersublattice exchange it
decreases. It is usual situation for physics of phase transitions,
because the growth of intrasublattice exchange at $I=0$ can result in
ferromagnetic state with characteristic to
such kind of transition the susceptibility singularity (it goes to the
infinity at the transition point). At the same time, the transition
to the antiferromagnetic state is not accompanied by the abnormal
growth of magnetic susceptibility. Indeed, the point of phase
transition to the antiferromagnetic phase corresponds the
equality $D=2(I+|J|)$. At its substitution in Eq. (32)
the value $\chi^{(0)}_{\perp}=1/2I$ is directly obtained. The same will be the value
of magnetic susceptibility in the antiferromagnetic phase, the
magnetization of which is determined by the expression (27).

     \section{Induced magnetostriction in the longitudinal magnetic field}

     Considering the striction properties of Van Vleck
     antiferromagnets, it will be  for certainty supposed, that
     the crystal has a hexagonal structure. There will be confined, at that, in
     magneto-elastic energy (see Eq. (1)) by spin-deformation
     interaction, that is proportional to second power of average
     spin [45]. Besides the single-ion terms will be accounted in the
     energy $E_{m-el}$; they contain the average
     components of spin quadrupole moment [46,47]. Then for
     such approaches the elastic and magneto-elastic contributions
     to the total
     energy (1) can be presented as following:

     \begin{eqnarray}
     E_{el}=\frac{1}{2}c_{11}\left(u^2_{xx}+u^2_{yy}\right)+\frac{1}{2}c_{33}u^2_{zz}+
     c_{12}u_{xx}u_{yy}+c_{13}\left(u_{xx}+u_{yy}\right)u_{zz}\nonumber\\+2c_{44}\left(
     u^2_{xz}+u^2_{yz}\right)+2c_{66}u^2_{xy}
     \end{eqnarray}
     \begin{eqnarray}
     E_{m-el}=\sum_{\alpha\beta}\left[ \lambda_{\alpha\beta}u_{zz}+
     \gamma_{\alpha\beta}\left(
     u_{xx}+u_{yy}\right)\right]\mathbf{s}_{\alpha}\mathbf{s}_{\beta}+
     \sum_{\alpha}\Bigl[B^{(s-i)}_{11}\left( Q^{XX}_{\alpha}u_{xx}
     +Q^{YY}_{\alpha}u_{yy}\right)\nonumber\\+B^{(s-i)}_{33}Q^{ZZ}_{\alpha}u_{zz}
     +B^{(s-i)}_{12}\left(Q^{XX}_{\alpha}u_{yy}+Q^{YY}_{\alpha}u_{xx}\right)+4B^{(s-i)}_{44}
     \left(Q^{YZ}_{\alpha}u_{yz}+Q^{XZ}_{\alpha}u_{xz}\right)\nonumber\\+
     4B^{(s-i)}_{66}Q^{XY}_{\alpha}u_{xy}\Bigr]
     +\sum_{\alpha\beta}\Bigl\lbrace
     B^{(\alpha\beta)}_{11}\left(s^X_{\alpha}s^X_{\beta}u_{xx}
     +s^Y_{\alpha}s^Y_{\beta}u_{yy}\right)+B^{(2)}_{33}s^Z_{\alpha}s^Z_{\beta}u_{zz}\\
     +B^{(\alpha\beta)}_{12}\left(
     s^X_{\alpha}s^X_{\beta}u_{yy}+s^Y_{\alpha}s^Y_{\beta}u_{xx}\right)+4B^{(\alpha\beta)}_{44}
     (s^Y_{\alpha}s^Z_{\beta}u_{yz}+s^X_{\alpha}s^Z_{\beta}u_{xz})
     +4B^{(\alpha\beta)}_{66}s^X_{\alpha}s^Y_{\beta}u_{xy}\Bigr\rbrace,\nonumber
     \end{eqnarray}
      where $\lambda_{\alpha\beta}$, $\gamma_{\alpha\beta}$ are
      the parameters of magneto-elastic exchange interactions, in
      which indices $\alpha$, $\beta$, as above, are the numbers of the spin
      sublattices, $B^{(s-i)}_{jl}$ and $B^{(\alpha\beta)}_{jl}$
      are the parameters of anisotropic magneto-elastic
      interactions [45], where the upper index shows on the
      single-ion or ionic origin, correspondingly; $u_{ij}$ are
      the components of elastic deformation tensor, $c_{jl}$ are
      the coefficients of elasticity. It should be noted, that single-ion
      magneto-elastic interactions in Eq. (34) are written in the
      crystallographic coordinate systems XYZ, so, as distinct from
      Eq. (5), the indices of quadrupole moment components $Q^{jl}=\frac{1}{2}
      \langle s^{j}s^{l}+s^ks^l\rangle$ are also defined in this
      system.

      The values of elastic deformations, which appear because of
      spin configuration change, will be found by the
      minimaization of energies (33) and (34) by corresponding
      components of deformation tensor. As a result the following
      expressions are obtained:

     \begin{eqnarray}
     u_{xx}+u_{yy}=-\frac{1}{c_{11}+c_{12}-2c^2_{13}/c_{33}}\Biggl[2\sum_{\alpha\beta}
     \gamma_{\alpha\beta}\mathbf{s}_{\alpha}\mathbf{s}_{\beta}
     +\sum_{\alpha}\left( B^{(s-i)}_{11}+B^{(s-i)}_{12}\right)\left( Q^{XX}_{\alpha}+Q^{YY}_{\alpha}\right)
    \nonumber\\+\sum_{\alpha\beta}\left(B^{(\alpha\beta)}_{11}+B^{(\alpha\beta)}_{12}\right)
     \left(s^X_{\alpha}s^X_{\beta}+s^Y_{\alpha}s^Y_{\beta}\right)-\frac{2c_{13}}{c_{33}}
    (\sum_{\alpha\beta}\lambda_{\alpha\beta}\mathbf{s}_{\alpha}\mathbf{s}_{\beta}
          +\sum_{\alpha}B^{(s-i)}_{33}Q^{ZZ}_{\alpha}\\+\sum_{\alpha\beta}
     B^{(\alpha\beta)}_{33}s^Z_{\alpha}s^Z_{\beta})\Biggr],\nonumber
     \end{eqnarray}
     \begin{eqnarray}
     u_{xx}-u_{yy}=-\frac{1}{c_{11}-c_{12}}\Bigl[\sum_{\alpha}\left( B^{(s-i)}_{11}-B^{(s-i)}_{12}\right)
     \left( Q^{XX}_{\alpha}-Q^{YY}_{\alpha}\right)+\sum_{\alpha\beta}
     \left(B^{(\alpha\beta)}_{11}-B^{(\alpha\beta)}_{12}\right)\times\nonumber\\\times
     \left(s^X_{\alpha}s^X_{\beta}-s^Y_{\alpha}s^Y_{\beta}\right)\Bigr],
     \end{eqnarray}
     \begin{eqnarray}
     u_{zz}=-\frac{\left(c_{11}+c_{12}\right)}{c_{33}\left(c_{11}+c_{12}\right)-2c^2_{13}}
     \Biggl[\sum_{\alpha\beta}\lambda_{\alpha\beta}\mathbf{s}_{\alpha}\mathbf{s}_{\beta}
     +\sum_{\alpha}B^{(s-i)}_{33}Q^{ZZ}_{\alpha}+\sum_{\alpha\beta}
     B^{(\alpha\beta)}_{33}s^Z_{\alpha}s^Z_{\beta}\nonumber\\-\frac{c_{13}}{c_{11}+c_{12}}
    \Bigl\{2\sum_{\alpha\beta}\gamma_{\alpha\beta}\mathbf{s}_{\alpha}\mathbf{s}_{\beta}
     +\sum_{\alpha}\left( B^{(s-i)}_{11}+B^{(s-i)}_{12}\right)
   \left( Q^{XX}_{\alpha}+Q^{YY}_{\alpha}\right)\\
    +\sum_{\alpha\beta}\left(B^{(\alpha\beta)}_{11}+B^{(\alpha\beta)}_{12}\right)
     \left(s^X_{\alpha}s^X_{\beta}+s^Y_{\alpha}s^Y_{\beta}\right)\Bigr\}\Biggr].\nonumber
     \end{eqnarray}

     The Eq. (35) determines the isotropic striction of
     "easy"$\,$ plane, or, what is the same, its expansion
     (or its contraction, depending on the signs of magneto-elastic
     coefficients), and Eq. (37) -- the expansion/contraction
     along the crystal symmetry axis.

     The spontaneous deformation in singlet phase is defined by obtained Eqs.
     (35)-(37) after substitution in them corresponding values of spin
     variables: $s=0$, $Q^{ZZ}=0$, $Q^{XX}=Q^{YY}=1$. After this it
     follows, that in this phase only the isotropic deformation of
     "easy"$\,$ plane and expansion/constriction will be not equal
     to zero:
     \begin{equation}
     u^{(0)}_{xx}+u^{(0)}_{yy}=-4\frac{B^{(s-i)}_{11}+B^{(s-i)}_{12}}{c_{11}+c_{12}-2c^2_{13}/c_{33}},
     \end{equation}
     \begin{equation}
      u^{(0)}_{zz}=4\frac{c_{13}\left(B^{(s-i)}_{11}+B^{(s-i)}_{12}\right)}{\left(c_{11}+c_{12}\right)c_{33}-2c^2_{13}},
     \end{equation}
      where index $(0)$ refers to the spontaneous
      magnetostriction. It can be seen, that in singlet phase the
      spontaneous deformations are defined only by single-ion
      magneto-elastic coefficients and satisfy the ratios:
      $u^{(0)}_{xx}=u^{(0)}_{yy}=-u^{(0)}_{zz}c_{33}/2c_{13}$. The
      expressions (38) and (39) remain correct in the magnetic
      field too, while $h_{\parallel}<h_s$, i.e. in the region of
      the singlet phase stability. In other words, the striction, that is
      specified in this region by Eqs. (38)-(39), does not depend
      on the field.

      The induced striction appears only after the
      spin polarization occurrence, or in the fields
      $h_{\parallel}>h_s$ [47], and is described by Eqs.
      (35)-(37).
      They were written in the general form and included all
      admitted phenomenological parameters of magneto-elastic
      coupling, that has both exchange (inter-ionic) and
      single-ion (because of the change of crystal field, that affects
      the
      ions) origin. So, at the analysis of magnetostriction, it should
      be considered several, by our opinion, interesting
      cases.

      At first, let consider the magnetostiction, that is caused
      by isotopic exchange interaction. In the most of magnets,
      the corresponding magneto-elastic coupling does not
      depend on spin directions and usually exceeds the
      anisotropic magneto-elastic one on more then order of
      magnitude.
      However, it is easy to convince, that despite the fact, that
      exchange magneto-elastic interaction does not depend on spin
      directions in crystal, the striction, that is generated
      by the external field, can be anisotropic.

      Indeed, it will be supposed, that in Eq. (34) only the
      magneto-elastic coefficients are finite $\lambda_{12}\neq0$,
      $\gamma_{11}\neq0$, and also $c_{13}\rightarrow 0$. Such a
      situation can take place, for example, in the lamellar
      crystals. If magnetic sublattices are formed by
      spins in basal planes, then the intersublattice
      antiferromagnetic exchange depends basically on inter-atomic
      distances along the crystal symmetry axis. As for
      intrasublattice one, it depends on the inter-ionic distances in this
      plane. Then from Eqs. (38) and (39) one can find, that in singlet phase
      all $u^{(0)}_{jj}=0$ and it is not influenced by the field.
      When spin polarization becomes finite, the
      exchange magnetostriction is represented by quite
      simple ratios:

       \begin{equation}
     \frac{u_{xx}}{u^{flip}_{xx}}=\frac{u_{yy}}{u^{flip}_{yy}}=\mathbf{s}^2=s^2
     \end{equation}
      \begin{equation}
     \frac{u_{zz}}{u^{flip}_{zz}}=\mathbf{s}_1\mathbf{s}_2=s^2\cos2\theta
     \end{equation}
     where $2\theta$ is, as above, the angle between sublattice
     spins,
     $u^{flip}_{xx}=u^{flip}_{yy}=-2\gamma_{11}/(c_{11}+c_{12})$ and
     $u^{flip}_{zz}=-2\lambda_{12}/c_{33}$ are the values of
     induced striction at $h_{\parallel}=h_{flip}$.$\;$ It should be noted,
     that for this case $u^{(0)}_{xx}=u^{(0)}_{yy}=u^{(0)}_{zz}=0$ also.

     It follows from Eqs. (40) and (41), that in the region near $h_{\parallel}\rightarrow h_s$
     of
     induced by external field phase transition
     there are singularities in striction field dependencies.
     The derivatives $\partial u_{zz}/\partial h_{\parallel}$ and
     $\partial u_{xx}/\partial h_{\parallel}$ will have a
     jump at this point.

     The field behavior of induced exchange striction, that is
     defined by Eqs. (40) and (41), is shown on Fig. 5. It can be
     seen, that  in the field $\mathbf{h}\parallel OZ$
     the exchange striction (it is normalized and -- depending
     on the sign of
     $\gamma_{11}$ -- can be both positive or negative), that is
     caused by the intrasublattice interaction, does not change
     its sign and only "follows"$\,$ the behavior
     $s^2(h_{\parallel})$. At the same time, the stiction, that is
     originated from intersublattice exchange (its sign depends on the sign of
     $\lambda_{12}$) for such a field orientation, is
     non-monotonic. This fact is a direct and simple consequence of change of
     $\cos2\theta$ sign. At first, the spin polarization grows
     (during the increasing of the field above $h_s$)
     practically at the opposite spin directions, so the striction
     (by absolute value) also increases. However, the further
     field growth gives rise to spin sublattice tilt,
     which becomes more and more noticeable, and in its own turn
     it causes the decreasing and passing through zero at $\theta=\pi/4$ of
      deformation.
      After such an angle configuration, when $\theta\rightarrow
      0$, the magnetostriction again increases, reaching the
      maximum in the field $h_{\parallel}=h_{flip}$.

       \begin{figure} \includegraphics[width=0.7 \textwidth]{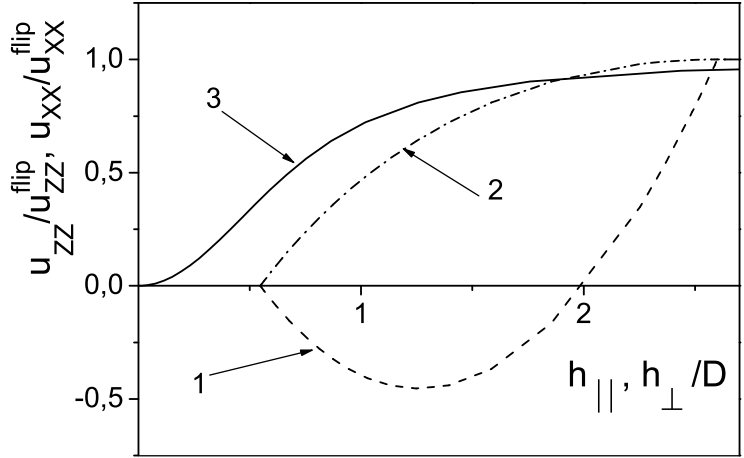}
       \caption{Exchange striction, which is described by Eqs.
       (40) and (41), for parameters $|J|/D=0.05$, $I/D=0.3$,
       $J_Z/D=1$. The line 1 corresponds to the "longitudinal"$\,$
      deformation, $u_{zz}/u^{flip}_{zz}$, and line 2 -- to the
       "transversal"$\,$ one $u_{xx}/u^{flip}_{xx}$ at $\mathbf{h}\parallel
       OZ$. The line 3 corresponds to the "longitudinal"$\,$
       deformation at $\mathbf{h}\perp OZ$.}
       \end{figure}

     According to Eqs. (40) and (41) the field behavior of
     $u_{xx}$ and $u_{zz}$ at $\mathbf{h}\perp
     OZ$ will be similar.  The exchange induced striction in this field
     changes monotonously, but begins to appear at the point $h_{\perp}=0$.
     It should be noted, that in the region $h_{\perp}\rightarrow
     0$ the striction is proportional to the $h^2_{\perp}$.
     Here also the derivatives $\partial u_{zz}/\partial h_{\perp}$
     and $\partial u_{xx}/\partial h_{\perp}$ changes continuously without jumps.

     Thus, the main distinction of induced
     striction at $\mathbf{h}\parallel OZ$ and $\mathbf{h}\perp
     OZ$ is, that in the longitudinal field
      there should be a jump in the field derivative of
     striction behavior versus field, and in the transversal
     field this derivative changes continuously.

     Let note, that field behavior of magnetostriction, shown on
     Fig. 5, is qualitatively agreed with experimental
     data, which are obtained from the measurement of DTN
     deformation [22,23]. Indeed, in this compound the
      intersublattice exchange dominates and resulting from it
      the magneto-elastic coupling refers to chains Ni-Cl-Ni-Cl,
      which are parallel to the axis $OZ$. It is still unknown,
      whether Ni
      ions, which lie in the one basal plane, form the one
      sublattice, because the nearest chains are shifted on the
      half of a period along axis $OZ$. But, nevertheless, there are no doubts,
      that in this singlet magnet the determinant (together with single-ion anisotropy)
       is the intersublattice (antiferromagnetic) exchange and its
       anisotropy.

      It is also not excluded, that in singlet magnets the
      anisotropy of magneto-elastic interaction can be comparable
      with isotropic one.
      The anisotropic part of magneto-elastic coupling may include
      both exchange (interion) and single-ion parts [4].
      Furthermore, it should be so in fact, if the magnetic
      system, to which DTN refers, is associated with strong
      single-ion anisotropy (which is the evidence of the
      essential
            spin-orbital interaction. In DTN, that is
      described by Hamiltonian (2), the single-ion
      anisotropy essentially exceeds the exchange interaction, the
      consequence of what is the formation of singlet ground
      state of the magnet in whole. So, in this case it is
      possible
      the next situation: the striction, that is caused by anisotropic
      interactions, exceeds the one, which is originated from isotropic
      exchange.

      As one more example, let consider such a case, when all
      magneto-elastic coefficients, except for $B^{(s-i)}_{33}$
      and $B^{(\alpha\beta)}_{33}$, can be neglected. Then, if
      also $c_{13}\rightarrow 0$, the main will be, as it is seen from
      Eq. (27), the deformation of crystal along axis $OZ$:

     \begin{eqnarray}
     u_{zz}=-\frac{2}{c_{33}}\left[B^{(s-i)}_{33}Q^{ZZ}_{\alpha}+\left(B^{(11)}_{33}+B^{(12)}_{33}\right)
     \left( s\cos\theta\right)^2\right].
     \end{eqnarray}
     This expression for striction can be written in the
     normalized form:
     \begin{equation}
     \frac{u_{zz}}{u^{flip}_{zz}}=\frac{B^{(s-i)}_{33}Q^{ZZ}_{\alpha}+\left(B^{(11)}_{33}
     +B^{(12)}_{33}\right)\left( s\cos\theta\right)^2}{B^{(s-i)}_{33}+
     B^{(11)}_{33}+B^{(12)}_{33}},
     \end{equation}
      where according to the definition
     \begin{eqnarray}
     u^{flip}_{zz}=-\frac{2}{c_{33}}\left(
     B^{(s-i)}_{33}+B^{(11)}_{33}+B^{(12)}_{33}\right)\nonumber
     \end{eqnarray}
     is the striction in the field $h_{\parallel}=h_{flip}$.

        \begin{figure} \includegraphics[width=0.7 \textwidth]{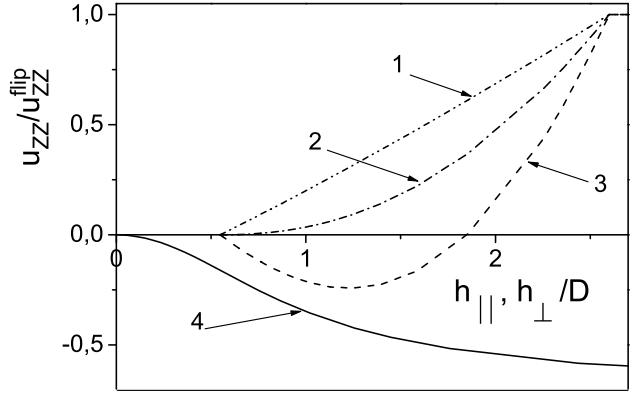}
        \caption{"Longitudinal"$\,$ $u_{zz}/u^{flip}_{zz}$
        magnetostriction \it versus \rm field. Line 1 is satisfied at $B^{(s-i)}_{33}\neq 0$
         and $B^{(11)}_{33}+B^{(12)}_{33}=0$, the line 2 --
         $B^{(s-i)}_{33}=0$, $B^{(11)}_{33}+B^{(12)}_{33}\neq 0$,
         the line 3 -- $\left(B^{(11)}_{33}+B^{(12)}_{33}\right)/
         \left(B^{(s-i)}_{33}+B^{(11)}_{33}+B^{(12)}_{33} \right)=2.5$,
          $B^{(s-i)}_{33}/\left(B^{(s-i)}_{33}+B^{(11)}_{33}+B^{(12)}_{33}\right)=-1.5$
         for $\mathbf{h}\parallel OZ$. The line 4 is obtained for the same parameter
         values as for line 3, but at $\mathbf{h}\perp OZ$}
        \end{figure}

 It is shown on Fig. 6 the field behavior of induced
 striction, which is obtained, using Eq. (43).
 The most interesting case is $\mathbf{h}\parallel OZ$, when
 $B^{(s-i)}_{33}\neq 0$, and $B^{(11)}_{33}+B^{(12)}_{33}=0$,
 to which refers the curve 1 on this figure. Corresponding
 magnetostriction is in direct proportion
 to the quadrupole moment $Q^{ZZ}$. It should be said, that obtained
 magneto-elastic contribution is linear (but not quadratic, as it is usually)
 by magnetization, taking into account, that,
 quadrupole moment as function of field is similar (see Figs. 1 and 3)
 to the behavior of $m(h_{\parallel})$.
 It should be also considered, that at the
 large exchange anisotropy $J_Z$ the magnetization
 versus $h_{\parallel}-h_s$ changes almost linearly, or
 $m_{\parallel}\sim h_{\parallel}-h_s$. So one can say, that the
 magnetostriction too (see line 1 on Fig. 6) behaves almost
 linearly in a whole region of antiferromagnetic phase existence.

 The field behavior of stiction at $\mathbf{h}\parallel OZ$,
 because of parameters choice, appears to be
 proportional to the $m^2_{\parallel}$, is corresponded with
 line 2 on Fig. 6.

 There is shown also on Fig. 6 the example (line 3) of common action
 of both anisotropic magneto-elastic striction mechanisms. At
 these ratios there takes place the competition between single-ion
 and inter-ion terms.
 In particular it is seen from line 3 on Fig. 6, that the field
 behavior of striction, that is caused by the anisotropic
 magneto-elastic interactions, can be the same as the stiction, that
 is originated from the isotropic exchange, what is shown on Fig. 5.
  However, it follows form Eq. (43), that at $\mathbf{h}\parallel OZ$
  the single-ion and inter-ion contributions in striction can compete, and
  at $\mathbf{h}\perp OZ$ there remains only the single-ion one.
  The striction behavior in $\mathbf{h}\perp OZ$ is shown as line
  4 on Fig. 6, at the same parameters, at which the line 2
  was obtained. Thus, in the considered examples of the competition between magneto-elastic
  interactions, it appeared, that in large fields
   $\mathbf{h}\parallel OZ$ and $\mathbf{h}\perp OZ$ the longitudinal striction
    has different signs.

  In DTN the components of tensor of longitudinal striction have one sign
  and are close in the values [22]. Therefore, the
    assumption, that the induced longitudinal striction
    at the magnetization in fields $\mathbf{h}\parallel OZ$ and
     $\mathbf{h}\perp OZ$, that are corresponded to
     $h_{\parallel,\perp}\rightarrow h_{flip}$ in this
    compound is caused by the intersublattice isotropic exchange
    interactions, is quite probable (believable). But, nevertheless, the additional
   investigations of magneto-elastic properties of the system are
    needed to prove unambiguously, that the observed deformation,
   under the effect of the field, is resulted from the interplane
   (in DTN -- intersublattice) exchange interaction only, and is not
   the the consequence of several field contributions, including from the
   spin quadrupole moment.

      \section {Conclusions}
     Thus, it was obtained, that the phase
     transition to magnetically ordered state, induced by magnetic
     field, in Van Vleck antiferromagnet, is the quantum phase
     transition.
       The spin polarization of the magnetic ion ground state
       is the order parameter of this phase
      transition, and for its description the Landau
      theory can be used. The considered transition is a
      consequence of competition of different interactions, and, what
      is important, it appears in the field, that is perpendicular
      to the easy plane. Such a field does not reduce the
      symmetry in this plane, leaving all directions in it
      equivalent. The conservation of degenaracy for sublattice
      magnetization directions in the easy plane is the crucial
      symmetrical condition of phase transition to the
      antiferromagnetic state with spontaneous magnetizations,
      that are lying in this plane. The account of stricition (see
      Eq. (36)) leads to the spontaneous lowering of the plane
      symmetry.

      It is also shown, that in the magnetic field induced
      antiferromagnetic phase the spin polarization
      (magnetization) of sublattice changes continuously from
      zero, reaching its maximum at the spin flipping point.
      As distinct from classical Neel antiferromagnets, in the magnetic ordered phase
      of Van Vleck (singlet) antiferromagnet the magnitude of
      sublattice magnetization strongly depends on the
      field. The same dependence on the field has an angle, that
      defines the deviation of sublattice magnetization from field direction.
      At the same time, the magnetization of a system as a whole,  being weekly dependent
      from the field, shows almost linear field
      behavior (when the exchange anisotropy is also accounted).

       The calculations indicate, that in such an antiferromagnet the
       induced magnetostriction appears only in the magnetic
       phase. This magnetostriction
       in the small fields region is connected with the
       spontaneous formation of sublattice magnetizations. In the large fields
       (which are corresponded to spin fliping field) the
       magnetostriction is basically determined by the sublattice magnetization rotation.

   Let emphasize two important issues. First one -- is the
   possibility of induced striction, which is originated by the intrasublattice
   magneto-elastic interaction.
    In the classical antiferromagnets this part of
   induced magnetostriction of antiferromagnets is usually
   neglected, because of the paraprocess smallness. The second
   aspect -- is connected with single-ion striction, the value of
   which is directly proportional to the spin quadrupole moment;
   as a result, it occurs that the striction, which is caused by
   this quantity, has close to linear dependence upon the field.

   Finally, the methodic notation should be made.
   The results, presented above, were obtained using the
   approximation of self-consistent field. It was supposed, that
   more accurate calculations will not bring any qualitative
   results, however they can influence quantitatively.
   Moreover, the magneto-elastic energy was written in the
   phenomenological form and contained a lot of parameters.
   So at the analysis of concrete compounds one should proceed from
   its characteristic hierarchy of interactions in the
   magneto-elastic energy, like it was made for example in this article.
   The separate paper will be devoted to detailed
   comparison of calculations with the available
   experimental data.

   We are grateful to Prof. S.M. Ryabchenko, who paid our
   attention on experimental articles [22,23] and the problem of
   magnetostricton properties of singlet magnets.


\begin{thebibliography}{99}
\bibitem{footnote1}
         As pointed in the recent paper [31], the nematic (in the
         very meaning of this term) ordering in strongly anisotropic
         magnets is absent. However the ground state of such magnet, that
         is singlet, do not defer, in physical sense, from previous one,
         that is caused by the large biquadratic exchange [27,28].
\bibitem{footnote2}
          It should be denoted, that ground state
          functions, given in [20] for laboratory (crystallographic) coordinates system,
          can not be considered as eigen ones, until the self-consistent problem
          is solved (see Ref. [38]).
\end{thebibliography}
      \end{document}